## Suppression of ferromagnetic order by Ag-doping: A neutron scattering investigation on $Ce_2(Pd_{1-x}Ag_x)_2In$ ($x$ = 0.20, 0.50)

A. Martinelli[1], M. Giovannini[2], J. G. Sereni[3], C. Ritter[4]

[1] *SPIN-CNR, C.so Perrone 24,I- 16152 Genova – Italy*

[2] *Dipartimento di Fisica, Universita` di Genova, Via Dodecaneso 33, I-16146 Genova, Italy*

[3] *Div. Bajas Temperaturas, Centro Atómico Bariloche (CNEA) and Conicet, 8400 S.C. Bariloche, Argentina*

[4] *Institut Laue - Langevin, 71 Avenue des Martyrs, F-38042 Grenoble Cedex 9, France*

### Abstract

The ground state magnetic behaviour of $Ce_2(Pd_{0.8}Ag_{0.2})_2In$ and $Ce_2(Pd_{0.5}Ag_{0.5})_2In$, found in the ferromagnetic branch of $Ce_2Pd_2In$, has been investigated by neutron powder diffraction at low temperature. $Ce_2(Pd_{0.8}Ag_{0.2})_2In$ is characterized by a ferromagnetic structure with the Ce moments aligned along the *c*-axis and values of 0.96(2) $\mu_B$. The compound retains the *P*4/*mbm* throughout the magnetic transition, although the magnetic ordering is accompanied by a significant decrease of the lattice strain along [00*l*], suggesting a magnetostructural contribution. The magnetic behaviour of $Ce_2(Pd_{0.5}Ag_{0.5})_2In$ is very different; this compound exhibits an extremely reduced magnetic scattering contribution in the diffraction pattern, that can be ascribed to a different kind of ferromagnetic ordering, with extremely reduced magnetic moments (~ 0.1 $\mu_B$) aligned along [0*l*0]. These results point to a competition between different types of magnetic correlations induced by Ag-substitution, giving rise to a magnetically frustrated scenario in $Ce_2(Pd_{0.5}Ag_{0.5})_2In$.

### 1. Introduction

Among the rare-earth-based intermetallic compounds, those with Ce and Yb exhibit unusual physical properties due to their *f*-electrons.[1,2,3] The members of the class of compounds with chemical formula $R_2T_2X$ (*R*: rare earth or U; *T*: Fe, Co, Ni, Ru, Rh, Pd; *X*: In, Sn, Pb) display a multiplicity of complex and interesting behaviours, depending on composition.[4,5,6] For instance, the $Ce_2Pd_2In$ compound, isotypic with $Mo_2FeB_2$, shows a significant range of solubility with two branches corresponding to a ferromagnetic (FM) and antiferromagnetic (AFM) arrangement of the magnetic moments, depending on the electron/hole concentration on the Pd/In layer.[7] In the AFM branch the hole concentration is increased by an excess of Pd substituting In vacancies, whereas in the FM branch the electronic concentration is driven by an excess of Ce. Since both magnetic structures, based on the competition between FM and AFM types of RKKY interactions, are comparable in energy, a large variety of magnetic behaviours have been observed in this family of alloys. Among different possibilities, the FM $Ce_2Pd_2In_1$ composition on the FM branch was studied by doping the 'hole-like' Pd lattice with 'electron-like' Ag.[8] In this series of alloys the FM character of their ground state, together with the low value of the paramagnetic temperature $\theta_P$ = -14 K and the observed magnetic moment $\mu_{eff}$ = 2.56 $\mu_B$/Ce_at, confirm the $Ce^{3+}$ character of the magnetic ion avoiding any relevant role of Kondo effect.

This change in the Coulomb potential sign on $Pd^+$ doped sites was shown to weaken the FM magnetic structure by changing FM interplane interaction into AFM through $Ag^+$ ions. As a result of the triangular Ce-Ce coordination in the basal plane, the growing AFM interactions leads to the formation of a frustrated phase maximized at $Ce_2(Pd_{0.5}Ag_{0.5})_2In_1$.[8] This phase is characterized by the rise of a specific heat ($C_m$) anomaly centred at $T_{max}$ ~1K, which increases with Ag concentration while the FM transition decreases in temperature and intensity of the $C_m$ jump till vanishing at the [Pd]/[Ag] = 1 concentration.

Whereas the behaviour of the intermediate $Ce_2(Pd_{1-x}Ag_x)_2In$ ($0 < x < 0.5$) compounds was described as due to the competition between a decreasing FM component versus an increasing frustrated component, the limit of solubility ($x$ = 0.5) was shown to fully fit into the systematic of frustrated systems.[9] Therefore, neutron scattering experiments, showing the evolution of the magnetic structure of these alloys, are expected to provide relevant information on the microscopic transference of degrees of freedom from the magnetically ordered to the frustrated configuration. This scenario provides a unique possibility to compare the evolution of ordered and frustrated configurations on the same system.

Only a few $Ce_2T_2X$ compounds have been investigated by neutron powder. In $Ce_2Ge_2In$ no evidence for magnetic Bragg scattering was observed down to 0.36 K.[10] For the composition $Ce_{2.22}Pd_{1.85}In_{0.78}$ the excess of Ce is located at the In 2*a* site;[5] moreover, a ferromagnetic ordering of the $Ce^{3+}$ moments takes place below $T_m$ ~ 3.8 K at both 4*h* and 2*a* sites. Conversely, in $Ce_{1.95}Pd_{2.35}In_{0.83}$ the excess of Pd partially occupies the 4*e* site;[5] remarkably, this composition displays an incommensurate antiferromagnetic ordering below $T_m$ ~ 4.0 K, similar to that observed in $Ce_2Pd_2Sn$.[11]

In this work we present a neutron scattering investigation on the alloys $Ce_{2.15}(Pd_{1-x}Ag_x)_{1.95}In_{0.9}$ ($x$ = 0.20 and 0.50) prepared in the FM branch. For simplicity's sake we will refer to them as $Ce_2(Pd_{1-x}Ag_x)_2In$.

## 2. Experimental

The two polycrystalline samples with nominal compositions $Ce_{2.15}(Pd_{1-x}Ag_x)_{1.95}In_{0.9}$ ($x$ = 0.20, 0.50) were prepared by argon arc melting the elements on a water cooled copper hearth with a tungsten electrode. The buttons were flipped over and remelted several times in order to ensure good homogeneity. The samples were then annealed at 750 °C for 168 h and finally water quenched.

The microstructure of the samples was observed after metallographic preparation by means of a scanning electron microscope (SEM; EVO 40 Carl Zeiss) equipped with an energy-dispersive X-ray spectroscope (EDS). Further details on starting materials, preparation and chemical characterization are reported in ref [8].

Neutron powder diffraction (NPD) analysis was carried out at the Institute Laue Langevin in Grenoble. Room temperature high-resolution data were collected at the D2B diffractometer ($\lambda$ = 1.5940 Å), whereas high-intensity data were collected at the D20 diffractometer ($\lambda$ = 2.4162 Å) between 7.0 and 1.45 K using an orange cryostat; for the sample with $x$ = 0.50, NPD data were also collected down to 100 mK using a $^3$He dilution cryostat at the same D20 diffractometer. Structural refinement was carried out according to the Rietveld method[12] using the program FULLPROF;[13] refinements were carried out using a file describing the instrumental resolution function. The microstructural evolution was investigated by refining the anisotropic strain parameters and analysing the broadening of diffraction lines by means of the Williamson–Hall plot method.[14] By means of representation analysis[15], possible spin orderings were evaluated, using the programs BASIREPS[16,17] and SARAh.[18]

## 3. Results

### *3.1 SEM analysis and structural properties*

SEM-EDS analysis reveals that the experimental stoichiometry of the investigated samples in is good agreement with the nominal one (Table 1). Trace amounts of a secondary phase with composition $Ce_{48}Ag_{52}$ have been observed in the $Ce_2(Pd_{0.5}Ag_{0.5})_2In$ sample.

Table 1: Chemical compositions of the investigated samples (EDS analysis).

| **Sample** | **Nominal composition** | **EDS (at%)** | | | |
|---|---|---|---|---|---|
| | | Ce | Pd | Ag | In |
| $Ce_2(Pd_{0.8}Ag_{0.2})_2In$ | $Ce_{42}Pd_{32}Ag_7In_{19}$ | 41.00 | 33.17 | 7.15 | 18.68 |
| $Ce_2(Pd_{0.5}Ag_{0.5})_2In$ | $Ce_{42}Pd_{22}Ag_{17}In_{19}$ | 41.74 | 21.91 | 17.40 | 18.95 |

From the structural point of view, the $R_2T_2X$ compounds crystallize in the tetragonal system and are isotypic with $Mo_2FeB_2$ (space group *P*4/*mbm*; number 127), an ordered derivative of the $U_3Si_2$-type structure. In particular, in the nominally stoichiometric $Ce_2Pd_2In$ the Ce, Pd and In atoms are located at the 4*h*, 4*g* and 2*a* sites, respectively, with Ce at the centre of a trigonal prism, coordinating 6 Pd atoms (Figure 1).

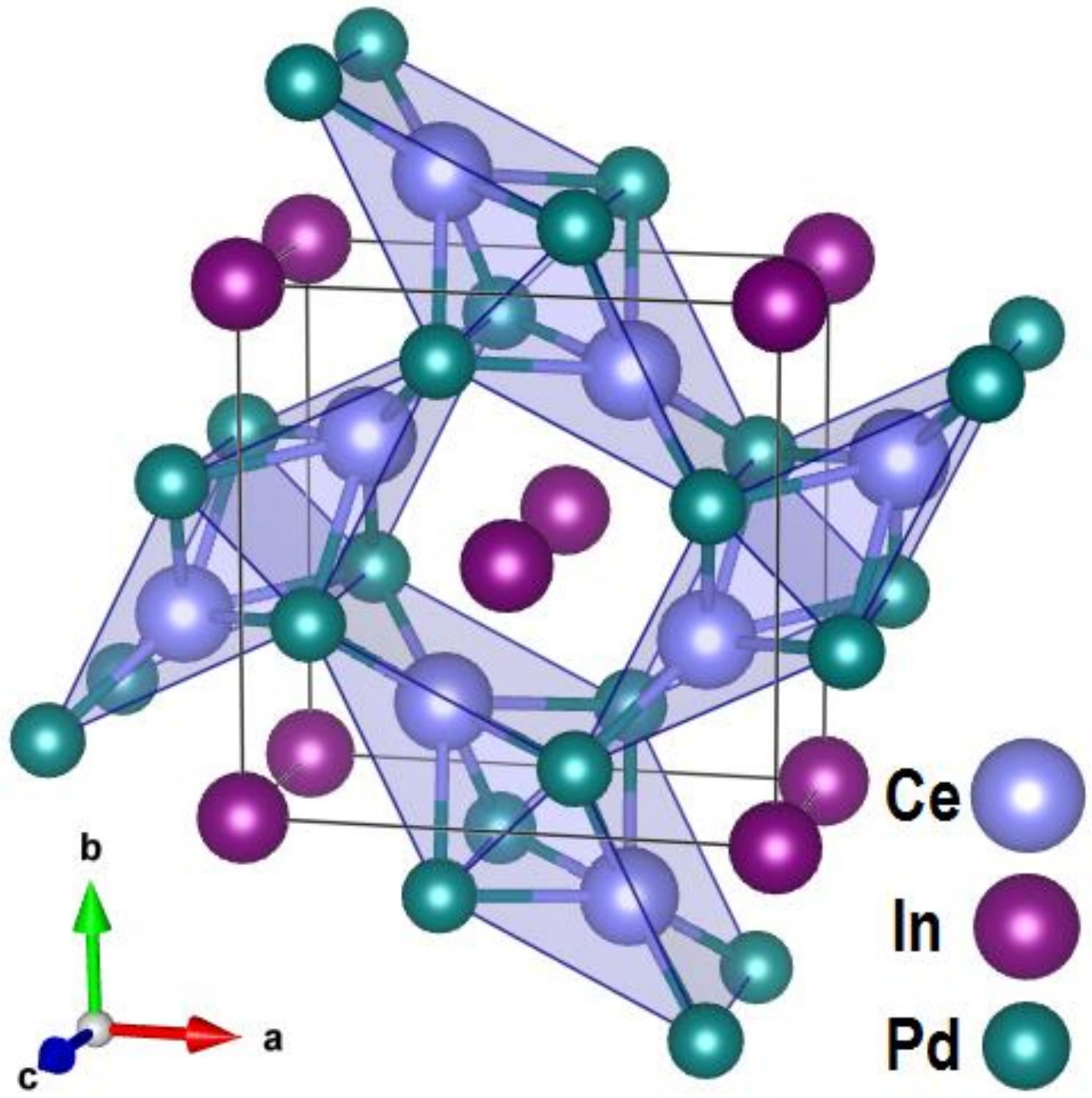

Figure 1: Crystal structure of $Ce_2(Pd_{1-x}Ag_x)_2In$.

### 3.2 NPD analysis of $Ce_2(Pd_{0.8}Ag_{0.2})_2In$

Rietveld refinements reveal that $Ce_2(Pd_{0.8}Ag_{0.2})_2In$ crystallizes in the tetragonal *P*4/*mbm* space group down to 1.45 K (Table 2 lists the structural data).

Table 2: Structural data of $Ce_2(Pd_{0.8}Ag_{0.2})_2In$ (space group: *P*4/*mbm*) as obtained by Rietveld refinement of NPD data collected at 300 K (D2B data) and 1.45 K (D20 data).

| | | **T = 300 K** | | | **T = 1.45 K** | | |
|---|---|---|---|---|---|---|---|
| *a* (Å) | | 7.8663(1) | | | 7.7929(1) | | |
| *c* (Å) | | 3.9185(1) | | | 3.9094(1) | | |
| Atomic site | | *x* | *y* | *z* | *x* | *y* | *z* |
| Ce | 4*h* | 0.1758(2) | 0.6758(2) | ½ | 0.1776(1) | 0.6776(1) | ½ |
| (Pd,Ag) | 4*g* | 0.3742(1) | 0.8742(1) | 0 | 0.3729(1) | 0.8729(1) | 0 |
| In | 2*a* | 0 | 0 | 0 | 0 | 0 | 0 |
| **m** Ce_1 ($\mu_B$) | | / | | | 0.96(2) | | |

| | | |
|---|---|---|
| $R_{Bragg}$ (%) | 8.17 | 6.77 |
| $R_F$ (%) | 4.30 | 3.82 |
| $R_{magnetic}$ (%) | / | 11.4 |

Below ~ 3.5 K neutron magnetic scattering is observed in the NPD data. The magnetic Bragg peaks can be successfully indexed according to the propagation vectors **k** = (0,0,0). As expected, magnetic ordering involves only the *R* atomic species (located at the 4*h* site) in the $R_2T_2X$ structure. By means of the representation analysis of magnetic structures,[15] the allowed magnetic moments **m** orderings at the 4*h* site were calculated. For a magnetic propagation vectors **k** = (0,0,0), the symmetry analysis using the representation theory provides the little group $\mathbf{G_k}$ = *P*4/*mbm*. Table 3 lists the resulting 10 irreducible representation (*irrep*) of $\mathbf{G_k}$; eight *irreps* are of dimension 1, whereas two *irreps* are of dimension 2. $\Gamma_m$ has the characters: $\chi(\Gamma_m - 4h)$ = (12,0,0,0,-2,-2,0,0,0,-4,0,0,-2,-2,0,0) and decomposes in terms of the *irreps* as $\Gamma_m(4h) = 1\,\Gamma_2 \oplus 1\,\Gamma_3 \oplus 1\,\Gamma_4 \oplus 1\,\Gamma_6 \oplus 1\,\Gamma_7 \oplus 1\,\Gamma_8 \oplus 1\,\Gamma_9 \oplus 2\,\Gamma_{10}$. This means that there is only 1 free parameter (the value of the magnetic moment) if a single $\Gamma_I$ describes the magnetic structure, where *I* = 2, 3, 4, 6, 7, 8. Conversely, $\Gamma_9$ and $\Gamma_{10}$ have dimension 2 and are present once and twice in $\Gamma_m(4h)$, respectively; hence, 2 and 4 parameters, respectively, are needed for their description (Table 4).

Table 3: Small irreducible representations of the little group $\mathbf{G_k}$ = *P*4/*mbm* for **k** = (0,0,0,) (symmetry operations are numbered according to the space group n° 127 of the International Tables for Crystallography;[19] the reader is deferred to this table for details).

| **Irrep** | **Symmetry operations** | | | | | | | | | | | | | | | |
|---|---|---|---|---|---|---|---|---|---|---|---|---|---|---|---|---|
| | 1 | 2 | 5 | 6 | 7 | 8 | 4 | 3 | 9 | 10 | 13 | 14 | 15 | 16 | 12 | 11 |
| $\Gamma_1$ | 1 | 1 | 1 | 1 | 1 | 1 | 1 | 1 | 1 | 1 | 1 | 1 | 1 | 1 | 1 | 1 |
| $\Gamma_2$ | 1 | 1 | 1 | 1 | 1 | 1 | 1 | 1 | -1 | -1 | -1 | -1 | -1 | -1 | -1 | -1 |
| $\Gamma_3$ | 1 | 1 | 1 | 1 | -1 | -1 | -1 | -1 | 1 | 1 | 1 | 1 | -1 | -1 | -1 | -1 |
| $\Gamma_4$ | 1 | 1 | 1 | 1 | -1 | -1 | -1 | -1 | -1 | -1 | -1 | -1 | 1 | 1 | 1 | 1 |
| $\Gamma_5$ | 1 | 1 | -1 | -1 | 1 | 1 | -1 | -1 | 1 | 1 | -1 | -1 | 1 | 1 | -1 | -1 |
| $\Gamma_6$ | 1 | 1 | -1 | -1 | 1 | 1 | -1 | -1 | -1 | -1 | 1 | 1 | -1 | -1 | 1 | 1 |
| $\Gamma_7$ | 1 | 1 | -1 | -1 | -1 | -1 | 1 | 1 | 1 | 1 | -1 | -1 | -1 | -1 | 1 | 1 |
| $\Gamma_8$ | 1 | 1 | -1 | -1 | -1 | -1 | 1 | 1 | -1 | -1 | 1 | 1 | 1 | 1 | -1 | -1 |
| $\Gamma_9$ | 1 0 | -1 0 | 1 0 | -1 0 | 0 1 | 0 -1 | 0 -1 | 0 1 | -1 0 | 1 0 | -1 0 | 1 0 | 0 -1 | 0 1 | 0 1 | 0 -1 |
| | 0 1 | 0 -1 | 0 -1 | 0 1 | 1 0 | -1 0 | 1 0 | -1 0 | 0 -1 | 0 1 | 0 1 | 0 -1 | 1 0 | 1 0 | -1 0 | 1 0 |
| $\Gamma_{10}$ | 1 0 | -1 0 | 1 0 | -1 0 | 0 1 | 0 -1 | 0 -1 | 0 1 | 1 0 | -1 0 | 1 0 | -1 0 | 0 1 | 0 -1 | 0 -1 | 0 1 |
| | 0 1 | 0 -1 | 0 -1 | 0 1 | 1 0 | -1 0 | 1 0 | -1 0 | 0 1 | 0 -1 | 0 -1 | 0 1 | -1 0 | -1 0 | 1 0 | -1 0 |

Table 4: Fourier coefficients obtained from the basis vectors for the different coordinates of the 4*h* site; *u*, *v*, *p*, *w* represent the free parameters of the possible magnetic structures.

| site 4*h* | | | $\Gamma_2$ | | | $\Gamma_3$ | | | $\Gamma_4$ | | | $\Gamma_6$ | | | $\Gamma_7$ | | | $\Gamma_8$ | | | $\Gamma_9$ | | | $\Gamma_{10}$ | | |
|---|---|---|---|---|---|---|---|---|---|---|---|---|---|---|---|---|---|---|---|---|---|---|---|---|---|---|
| *x* | *y* | *z* | *x* | *y* | *z* | *x* | *y* | *z* | *x* | *y* | *z* | *x* | *y* | *z* | *x* | *y* | *z* | *x* | *y* | *z* | *x* | *y* | *z* | *x* | *y* | *z* |
| *x* | *x*+½ | ½ | u | u | 0 | 0 | 0 | u | u | -u | 0 | u | u | 0 | 0 | 0 | u | u | -u | 0 | 0 | 0 | u-v | u+p | v+w | 0 |
| $\bar{x}$ | $\bar{x}$ +½ | ½ | -u | -u | 0 | 0 | 0 | u | -u | u | 0 | -u | -u | 0 | 0 | 0 | u | -u | u | 0 | 0 | 0 | -u+v | u+p | v+w | 0 |
| $\bar{x}$ +½ | *x* | ½ | -u | u | 0 | 0 | 0 | -u | -u | -u | 0 | u | -u | 0 | 0 | 0 | u | u | u | 0 | 0 | 0 | -u-v | p-u | v-w | 0 |
| *x*+½ | $\bar{x}$ | ½ | u | -u | 0 | 0 | 0 | -u | u | u | 0 | -u | u | 0 | 0 | 0 | u | -u | -u | 0 | 0 | 0 | u+v | p-u | v-w | 0 |

Figure 2 shows the Rietveld refinement plot obtained with NPD data collected at 1.45 K, whereas the inset illustrates the superposition of the data collected at 7.0 K and 1.45 K in the Q range where main magnetic peaks are observed; the ferromagnetic structure is illustrated in Figure 3.

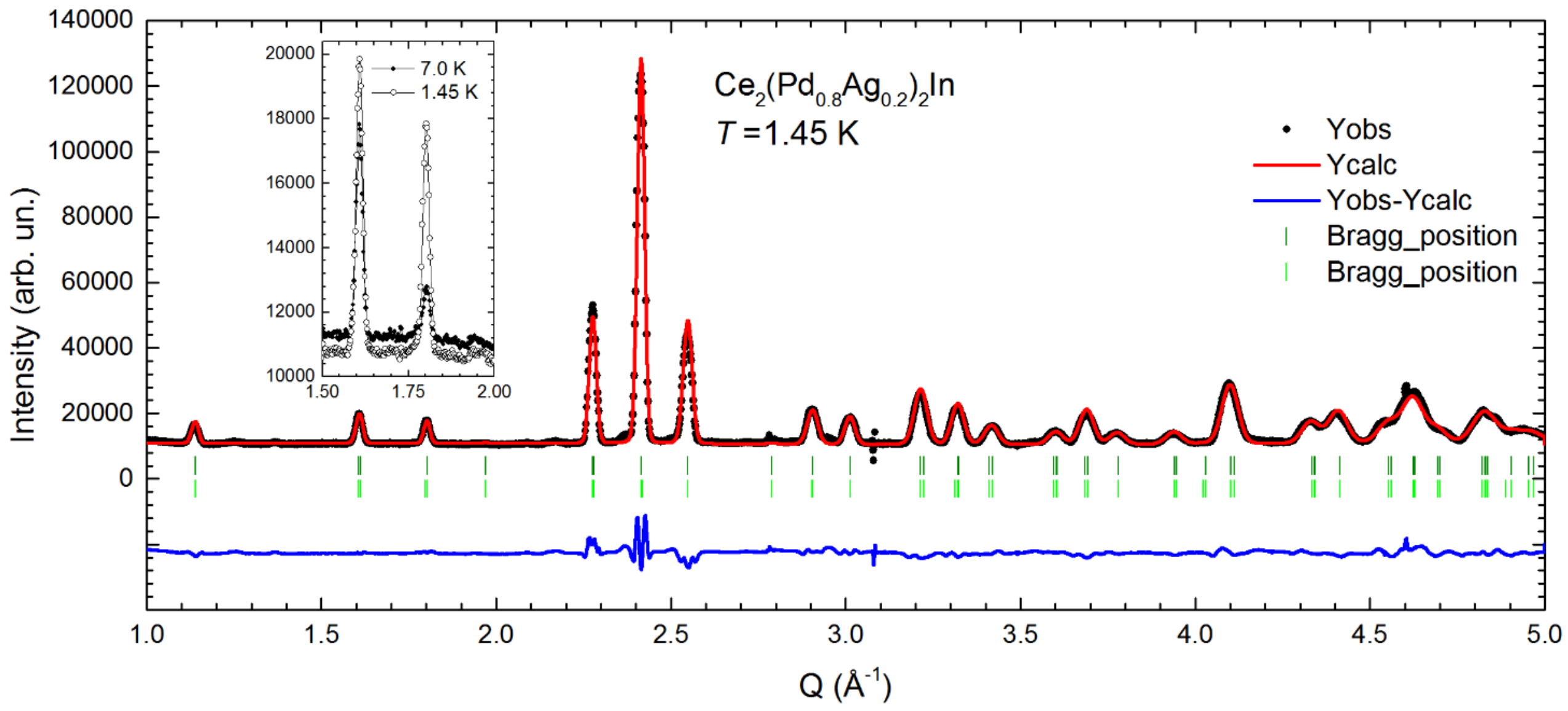

Figure 2: Rietveld refinement plot for $Ce_2(Pd_{0.8}Ag_{0.2})_2In$ (NPD data collected at D20); in the upper field the continuous red line is the calculated fit superposed to the experimental intensity data (black points); the blue line in the lower field is the difference curve; the green vertical bars indicate Bragg reflections for the crystallographic and magnetic phases. The inset shows the superposition of the data collected at 7.0 K and 1.45 K in the Q range where main magnetic peaks occur.

The fit of the experimental data has been obtained by applying the *irrep* $\Gamma_7$ for $\mathbf{G_k}$, corresponding to a ferromagnetic ordering at the Ce (4*h*) sub-lattice with magnetic moments $\mathbf{m}$ = 0.96(2)$\mu_B$ parallel to the tetragonal *c*-axis. This magnetic structure belongs to the Shubnikov magnetic space group *P*4/*mb*'*m*' (magnetic space group number: 127.393). Remarkably, the refined magnetic moment value is notably reduced in comparison with the saturation moment of the free $Ce^{3+}$ ion (2.54 $\mu_B$) expected for the six-fold Hund's rules ground state (GS) for a total angular momentum $J$ = 5/2. At low temperatures (i.e. 1.45 and 7 K) only a doublet GS is expected to contribute to the magnetic signal due to crystal field effects and Kondo interactions. From the low paramagnetic temperature values $\theta_P$ ~ -10 K no significant Kondo effects are expected in this system.[5]

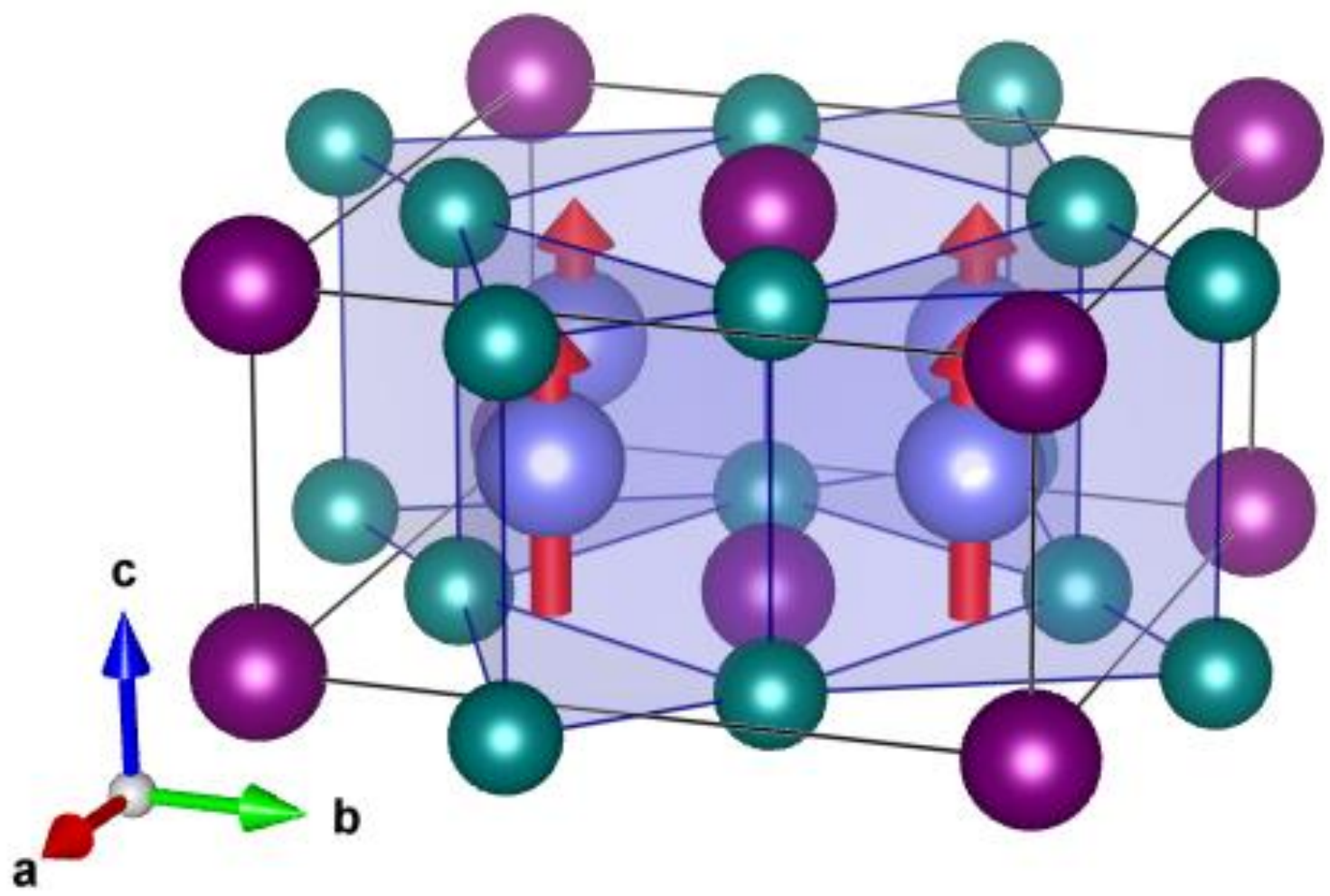

Figure 3: Magnetic moments arrangement at the Ce sub-lattice in $Ce_2(Pd_{0.8}Ag_{0.2})_2In$ at 1.45 K.

The thermal dependence of the ordered magnetic moment (Figure 4) which represents the Landau order parameter in a FM system, agrees with the magnetization curve obtained from DC susceptibility measurements.[8] These features are consistent with a 2nd order nature of the magnetic transition shown by the specific heat jump at $T_C$ = 3.3K. The Landau fit of the thermal dependence of the ordered magnetic moment extrapolates at the same temperature (see Figure 4).

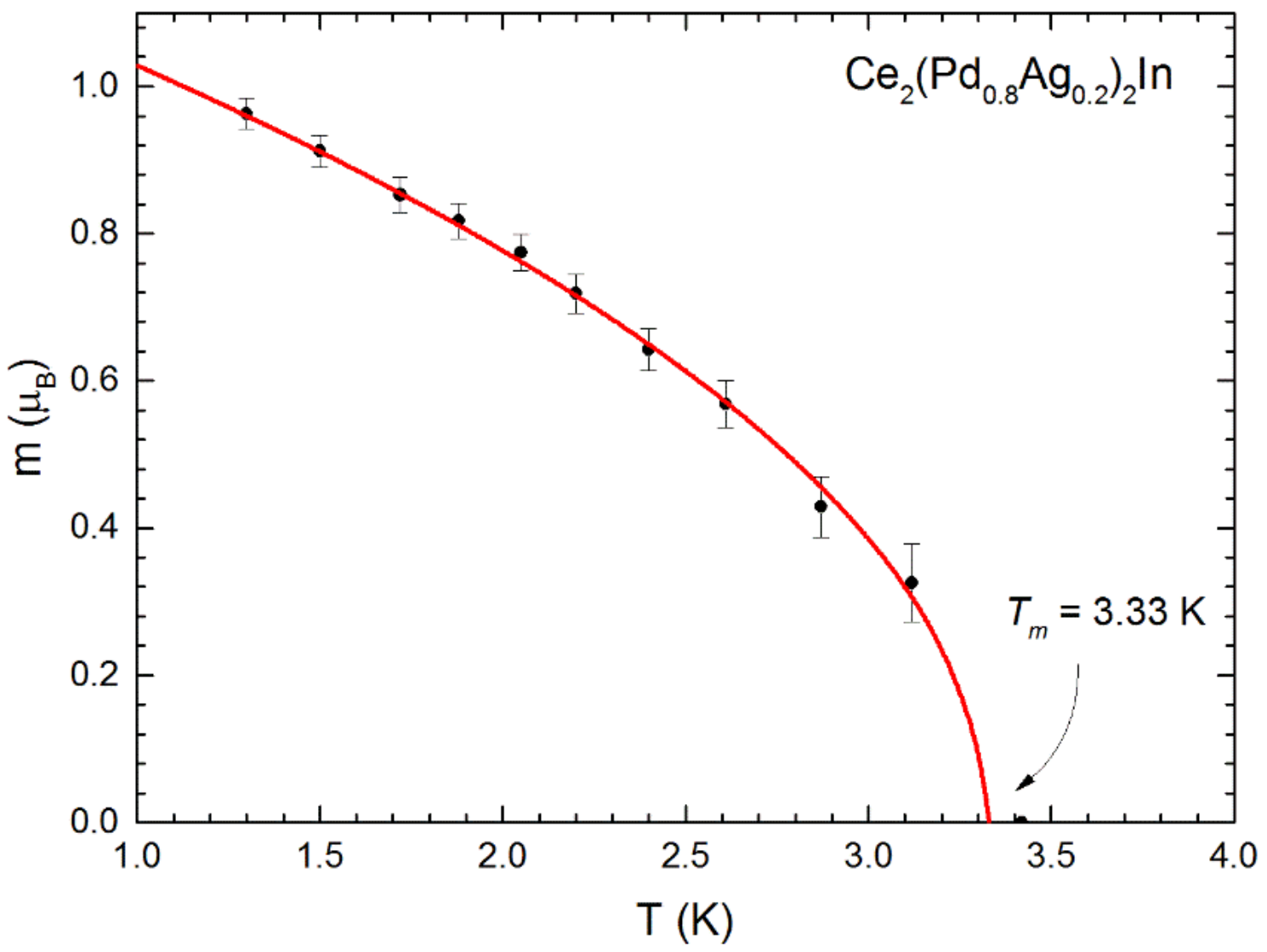

Figure 4: Thermal dependence of the ordered magnetic moment at the Ce sub-lattice in $Ce_2(Pd_{0.8}Ag_{0.2})_2In$; the solid line represents the Landau fit for a 2nd order transition.[20]

The ferromagnetic ordering observed in $Ce_2(Pd_{0.8}Ag_{0.2})_2In$ is similar to that observed in $Ce_2Pd_2(In_{0.6}Ce_{0.4})$, where the exceeding Ce atoms are located at the In 2*a* site.[5] The main difference is that in $Ce_2Pd_2(In_{0.6}Ce_{0.4})$ the magnetic moments of Ce order at both 4*h* and 2*a* sub-lattices, whereas in $Ce_2(Pd_{0.8}Ag_{0.2})_2In$ the amount of Ce at the 2*a* site is extremely low and hence its coherent magnetic scattering contribution, if present, is negligible.

The occurrence of the magnetic ordering is accompanied by a remarkable decrease of the lattice strain along [00*l*], as evidenced by the net decrease on cooling of the slope of the straight line connecting the 00*l* data in the Williamson-Hall plots reported in Figure 5.

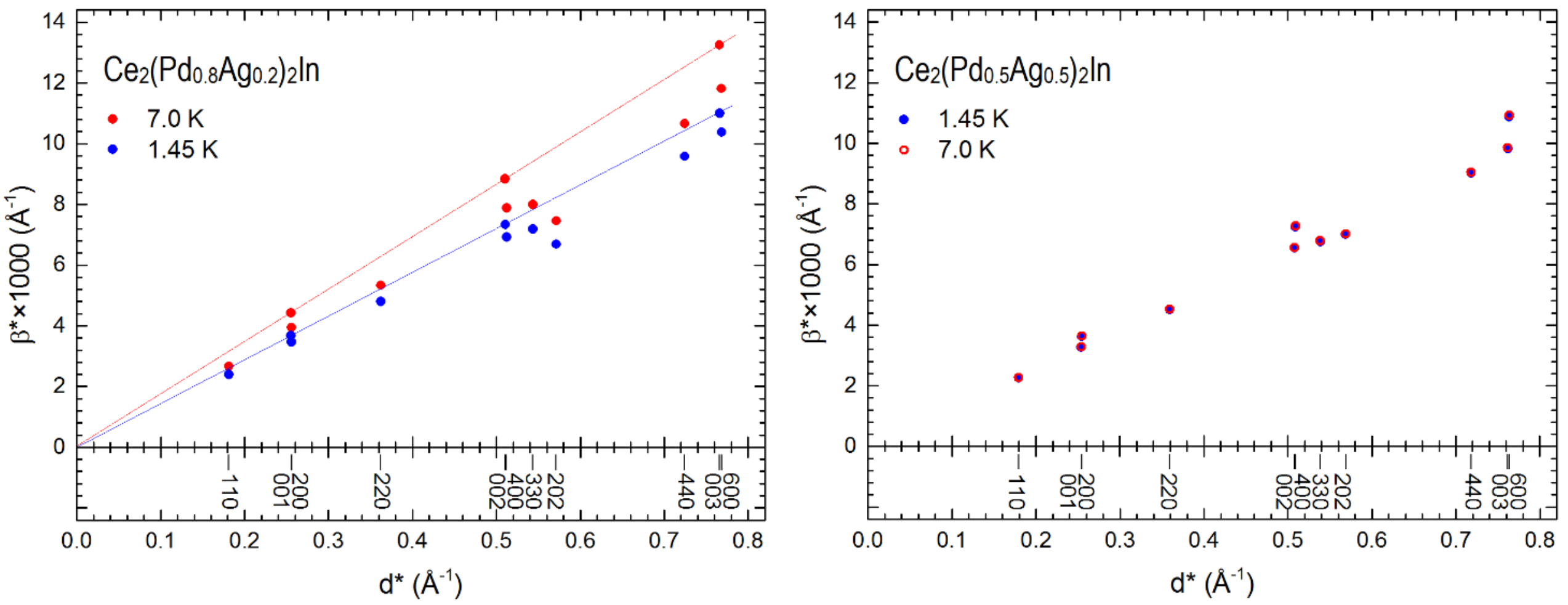

Figure 5: Superposition of the Williamson-Hall plots (data from selected families of crystallographic planes with different orders) obtained using the NPD data collected at 7.0 K and 1.45 K for $Ce_2(Pd_{0.8}Ag_{0.2})_2In$ (on the left) and $Ce_2(Pd_{0.5}Ag_{0.5})_2In$ (on the right); in the lower field the corresponding indexed Bragg peaks are listed. The straight lines are a guide to the eye connecting the 00*l* reflections.

### 3.3 NPD analysis of $Ce_2(Pd_{0.5}Ag_{0.5})_2In$

Figure 6 shows the Rietveld refinement plot obtained using NPD data collected at room temperature. Rietveld refinement reveals that the secondary phase detected by SEM-EDS analysis is constituted of cubic CeAg (space group $Pm\bar{3}m$); in particular, this secondary phase constitutes about 10-15% wt of the sample.

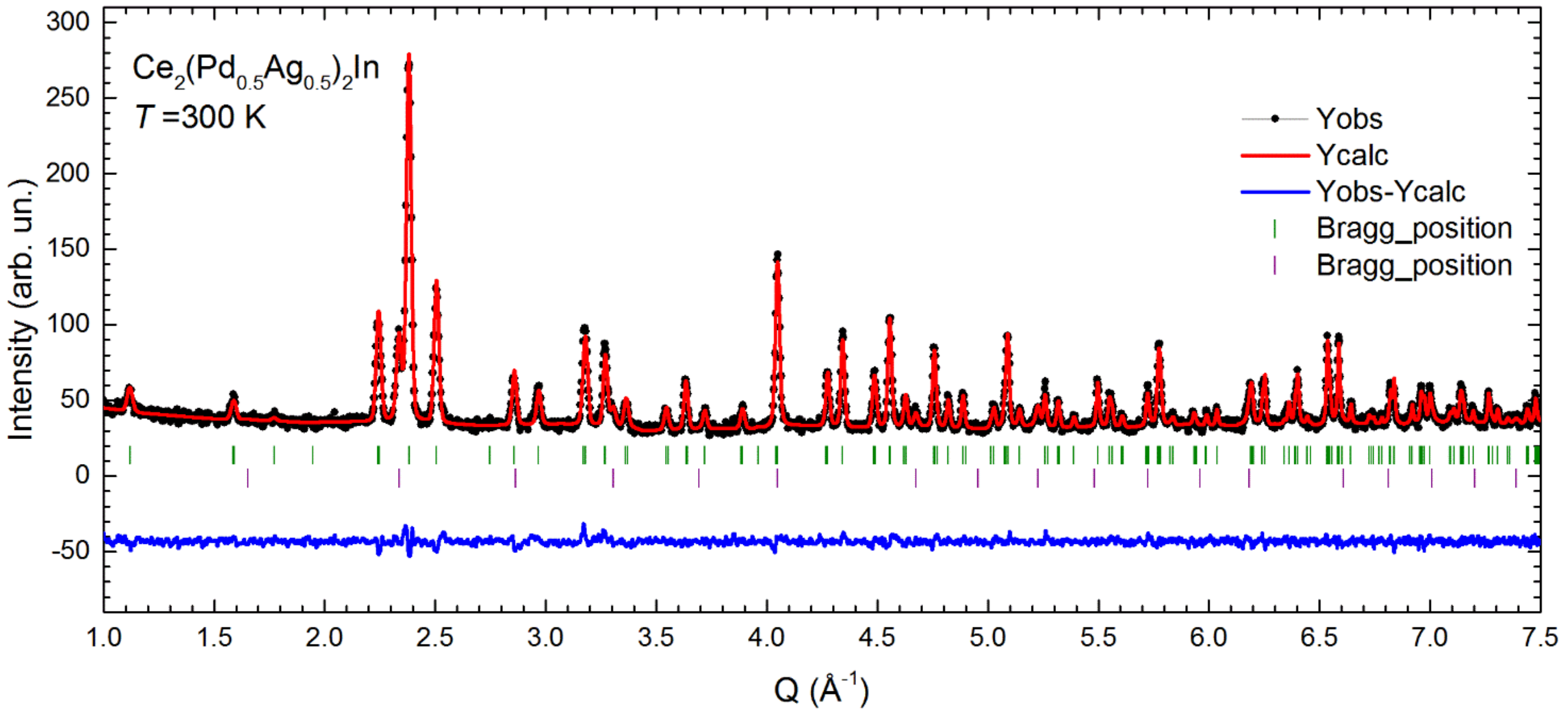

Figure 6: Rietveld refinement plot for $Ce_2(Pd_{0.5}Ag_{0.5})_2In$ (NPD data collected at D2B); in the upper field the continuous red line is the calculated fit superposed to the experimental intensity data (black points); the blue line in the lower field is the difference curve; the vertical bars indicate Bragg reflections for the $Ce_2(Pd_{0.5}Ag_{0.5})_2In$ (green) and CeAg (purple) phases.

Also for this composition, no structural transformation can be detected in the inspected thermal range (Table 5 lists the structural data at room temperature and 1.45 K).

Table 5: Structural data of $Ce_2(Pd_{0.5}Ag_{0.5})_2In$ (space group: *P*4/*mbm*) as obtained by Rietveld refinement of NPD data collected at 300 K (D2B data) and 1.45 K (D20 data).

| | | **T = 300 K** | | | **T = 1.45 K** | | |
|---|---|---|---|---|---|---|---|
| *a* (Å) | | | 7.9273(1) | | | 7.8772(1) | |
| *c* (Å) | | | 3.9505(1) | | | 3.9291(1) | |
| Atomic site | | *x* | *y* | *z* | *x* | *y* | *z* |
| Ce | 4*h* | 0.1767(2) | 0.6767(2) | ½ | 0.1777(1) | 0.6777(1) | ½ |
| (Pd,Ag) | 4*g* | 0.3752(2) | 0.8752(2) | 0 | 0.3737(1) | 0.8736(1) | 0 |
| In1 | 2*a* | 0 | 0 | 0 | 0 | 0 | 0 |
| $R_{Bragg}$ (%) | | | 4.99 | | | 8.61 | |
| $R_F$ (%) | | | 3.63 | | | 5.20 | |

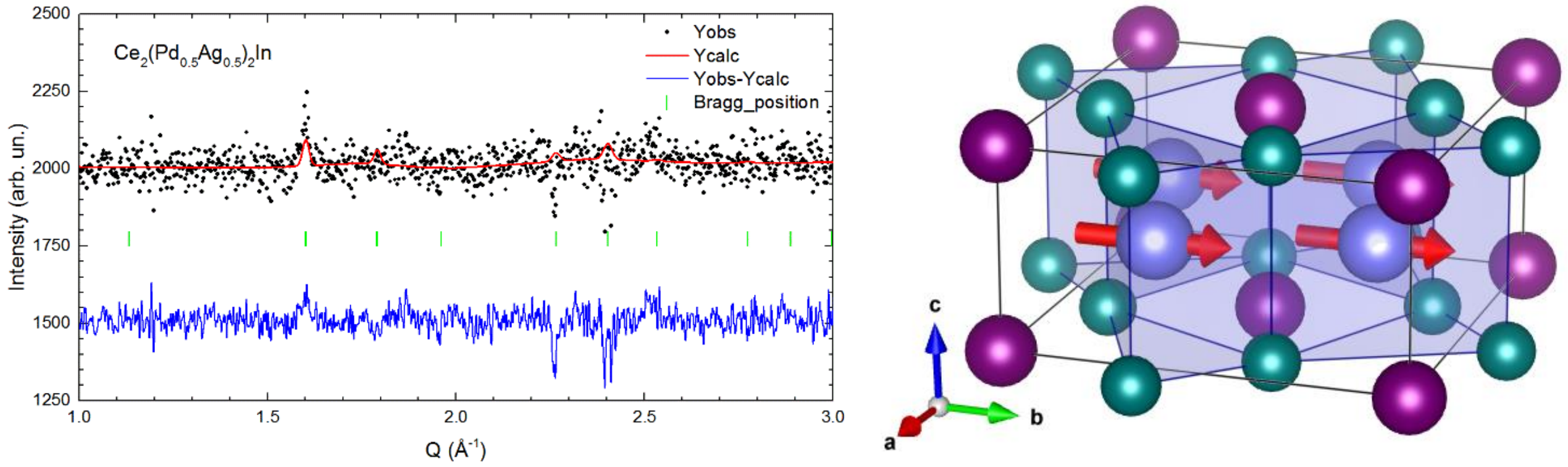

Figure 7: on the left: Fitted difference pattern for $Ce_2(Pd_{0.5}Ag_{0.5})_2In$ (1.45 K - 1.6 K data; orange cryostat data); on the right: magnetic structure corresponding to the *irrep* $\Gamma_{10}$.

Data collected using the orange cryostat reveal that the magnetic ordering corresponding to the *irrep* $\Gamma_7$ observed in $Ce_2(Pd_{0.8}Ag_{0.2})_2In$ is completely suppressed in $Ce_2(Pd_{0.5}Ag_{0.5})_2In$. A closer inspection reveals that NPD data collected at 1.6 K evidence no significant coherent magnetic scattering, whereas at 1.45 K a very faint magnetic contribution can be detected at **Q** ~1.6 Å$^{-1}$. Despite the paucity of the data, we tentatively succeed in determining a possible magnetic ordering at the Ce sub-lattice consistent with the *irrep* $\Gamma_{10}$ (Figure 7, on the left); an in-plane ferromagnetic ordering results (Figure 7, on the right), with extremely an reduced magnetic moments **m** ~ 0.12(1) $\mu_B$ aligned along the *b*-axis. Actually, the extremely reduced value of the ordered magnetic moment indicates that this is not a classical ferromagnetic state, but rather that a very faint resultant emerges from a disordered arrangements of the magnetic moments. This scenario is confirmed by the data subsequently collected down to 0.1 K using the $^3$He dilution cryostat, as illustrated in Figure 8 showing the progressive increase of the magnetic 001 peak located at **Q** ~1.6 Å$^{-1}$.

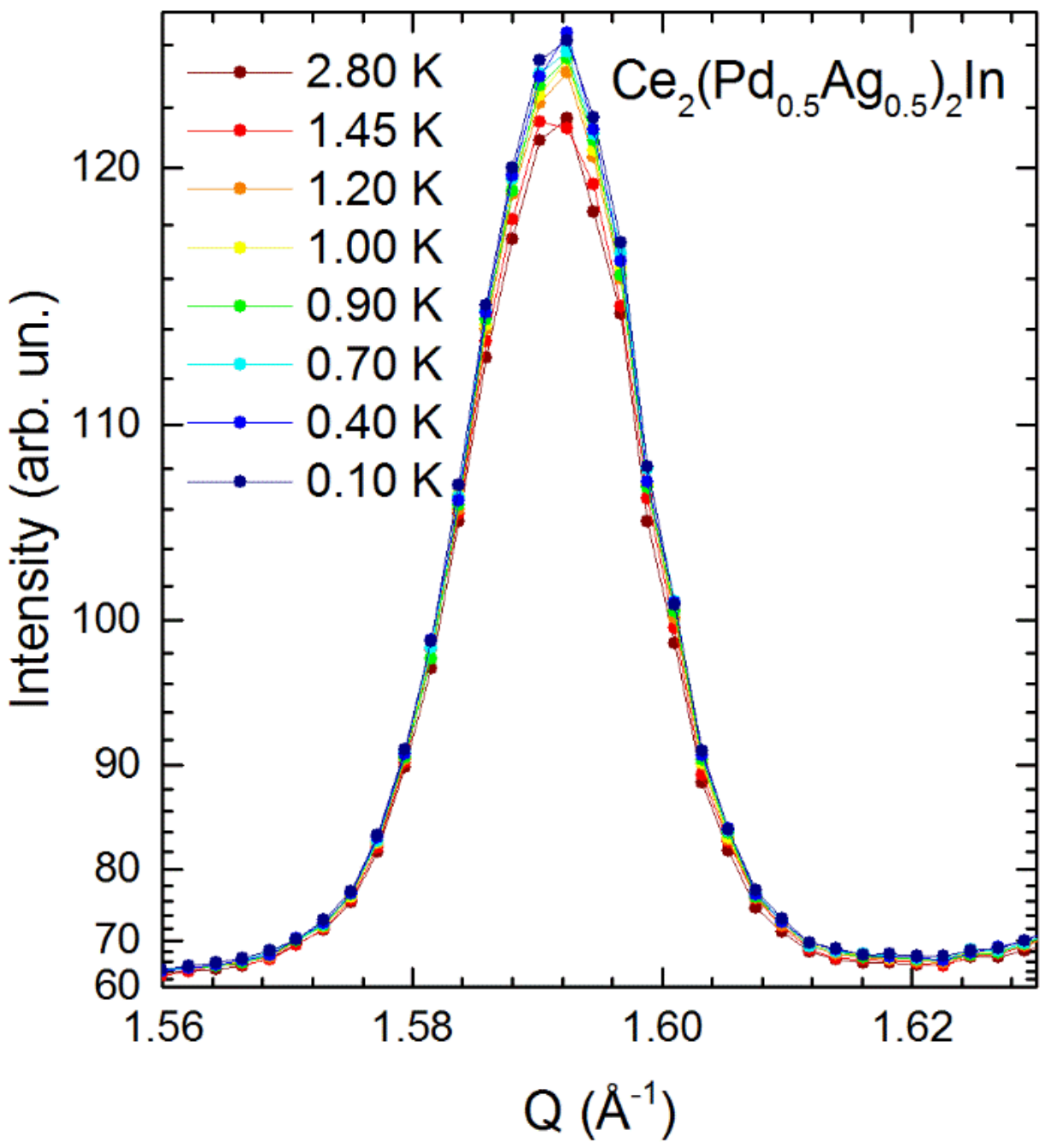

Figure 8: Thermal evolution of the coherent magnetic scattering contribution around **Q** ~1.6 Å$^{-1}$ ($^3$He dilution cryostat data; data intensity axis is not linearly scaled in order to highlight the faint increase originated by the magnetic scattering).

Regrettably, the statistic of these data is strongly decreased on account of the experimental set-up. In addition, the frustration characterizing this composition yields an extremely reduced coherent scattering contribution to the NPD pattern and the refined magnetic moment at 0.1 K is **m** ~ 0.17(6) $\mu_B$, consistent with the results obtained with orange cryostat at 1.45 K.

The analysis of the Williamson-Hall plots collected at 7 K and 1.45 K (Figure 5) reveals that the lattice strain is constant in this thermal range, conversely to what observed in $Ce_2(Pd_{0.8}Ag_{0.2})_2In$, where the rearrangement of the magnetic moments determines a significant increase of the strain along [001].

## 4. Discussion

Figure 9 shows how the crystal structure of $Ce_2(Pd_{0.8}Ag_{0.2})_2In$ is distorted by increasing the Ag content up to the $Ce_2(Pd_{0.5}Ag_{0.5})_2In$ composition. The Ag plane are displaced in the plane and the resulting structural distortion induces a displacement of the Ce atoms towards the (Pd,Ag) positions themselves, determining an increase of the shortest Ce-Ce bond distance (the Ce(1)-Ce(5) bond length in Table 6 below).

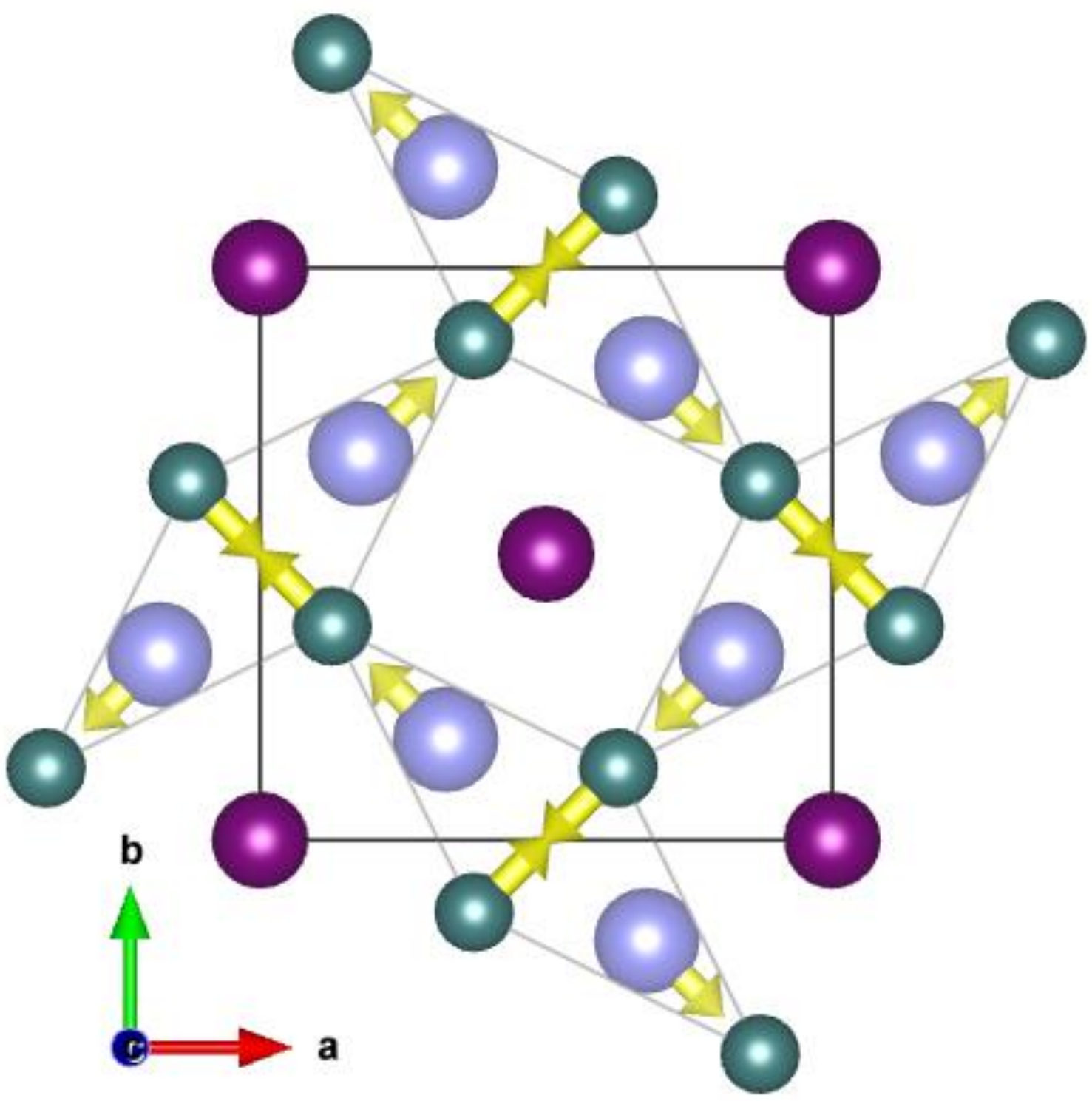

Figure 9: Displacement of atoms observed in $Ce_2(Pd_{0.5}Ag_{0.5})_2In$ compared to the reference $Ce_2(Pd_{0.8}Ag_{0.2})_2In$ structure (structural data at 300 K; atomic displacements are arrowed).

Table 6: Selected bond length in $Ce_2(Pd_{0.8}Ag_{0.2})_2In$ and $Ce_2(Pd_{0.5}Ag_{0.5})_2In$ at 7.0 K and 1.45 K; atoms numbering is depicted in Figure 10.

| | $Ce_2(Pd_{0.8}Ag_{0.2})_2In$ | | $Ce_2(Pd_{0.5}Ag_{0.5})_2In$ | |
|---|---|---|---|---|
| bond | 1.45 K<br>length (Å) | 7.0 K<br>length (Å) | 1.45 K<br>length (Å) | 7.0 K<br>length (Å) |
| Ce(1)-Ce(2,3,4,6) | 4.0567(2) | 4.0697(5) | 4.1059(5) | 4.1042(4) |
| Ce(1)-Ce(5) | 3.9142(2) | 3.8933(5) | 3.9247(5) | 3.9371(4) |
| Ce(1)-Ce(7,8) | 3.9094(1) | 3.9168(1) | 3.9308(1) | 3.9300(1) |
| Ce(1)-Pd(1,4) | 2.9078(1) | 2.9258(3) | 2.9529(3) | 2.9484(3) |
| Ce(1)-Pd(2,3,5,6) | 3.1004(1 | 3.0961(4) | 3.1119(3) | 3.1151(3) |
| Ce(1)-In(1,2,3,4) | 3.4712(1) | 3.4809(3) | 3.5062(3) | 3.5049(2) |

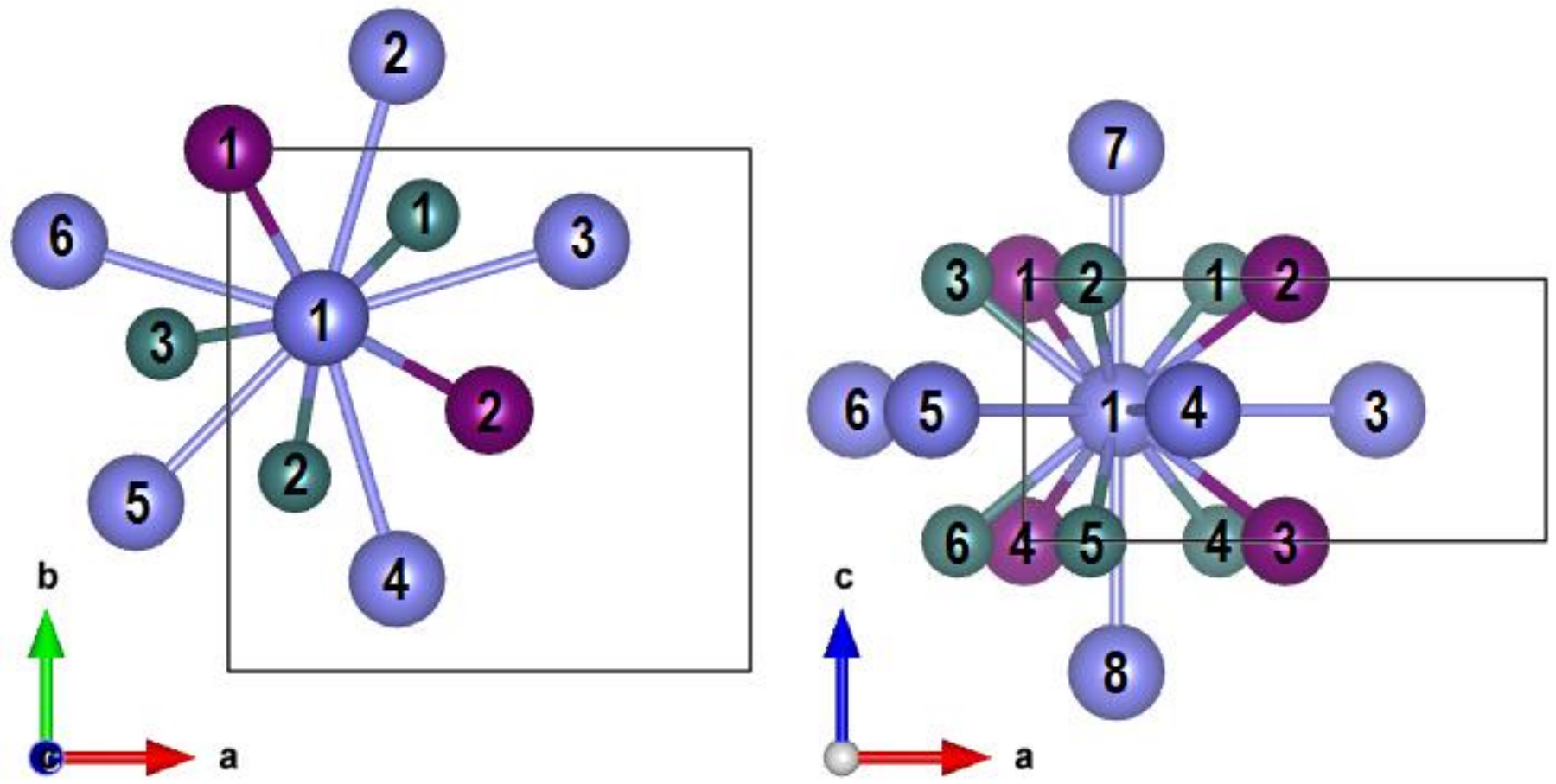

Figure 10: Coordination around the Ce atom; atoms are numbered as in Table 6.

Useful insights can be gained by comparing the crystal structures of both compounds at 7.0 K and 1.45 K, that is above and below the magnetic transition; Table 6 lists the bond lengths at 7.0 K and 1.45 K as described in Figure 10, whereas the atomic displacements are depicted in Figure 11. In both compounds the Ce(1)-Ce(5) bond length exhibits an anomalous change, undergoing a remarkable elongation in $Ce_2(Pd_{0.8}Ag_{0.2})_2In$ (~ 0.014 Å) and a likewise sizeable shrinking in $Ce_2(Pd_{0.5}Ag_{0.5})_2In$ (~ 0.012 Å). Likely, this different behaviour is originated by the different magnetic interactions at play in these compounds.

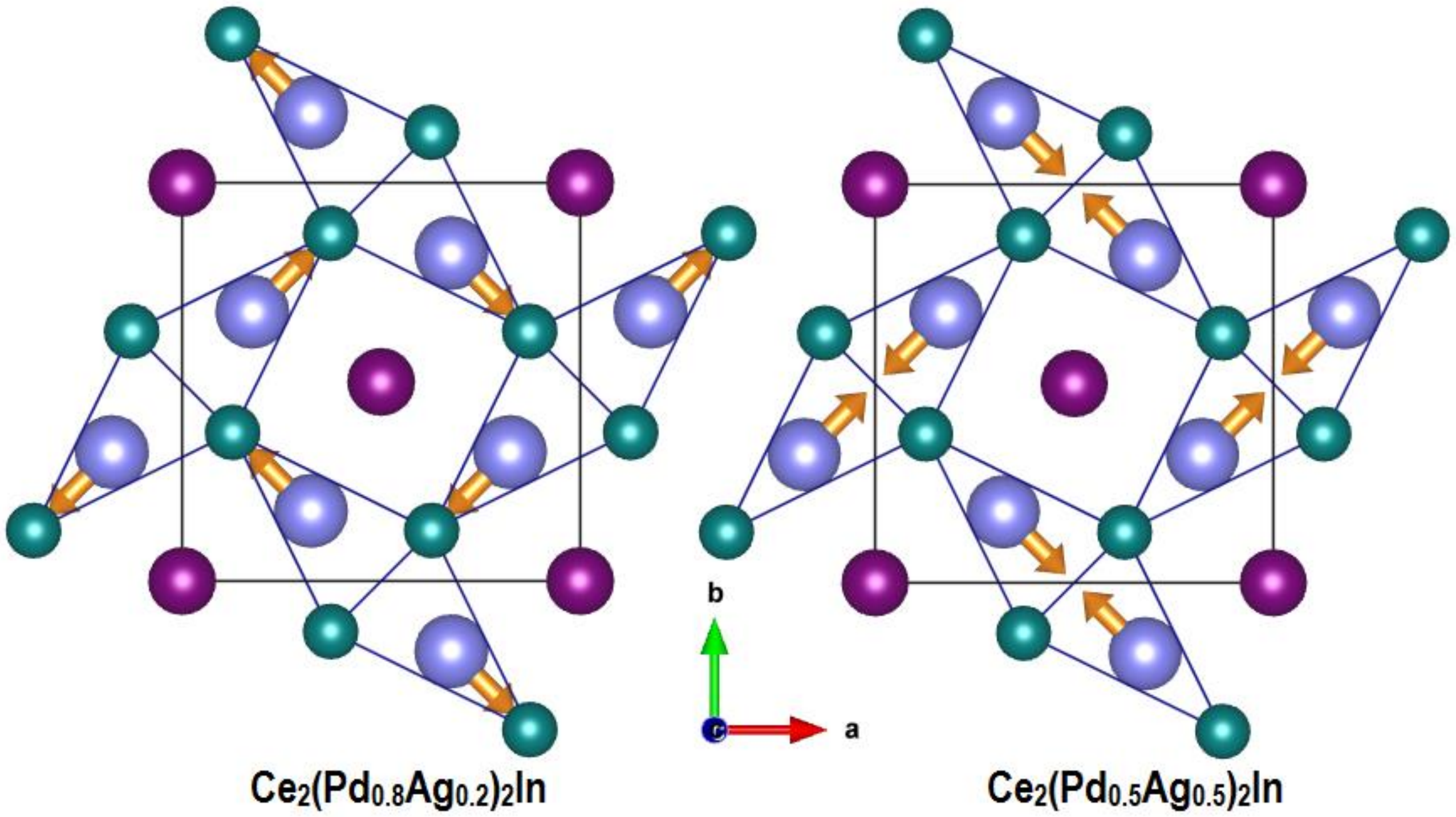

Figure 11: Structural distortion in the $Ce_2Pd_2(In_{0.6}Ce_{0.4})$ and $Ce_2(Pd_{0.8}Ag_{0.2})_2In$ compounds throughout the magnetic transition; atomic displacements are arrowed.

The magnetic and thermal properties of $Ce_2(Pd_{0.8}Ag_{0.2})_2In$ and $Ce_2(Pd_{0.5}Ag_{0.5})_2In$ were previously investigated in ref. [8] by magnetization measurements and specific heat analysis. Figure 12 compares specific heat and ac-susceptibility bulk measurements; complementary results on dc-magnetization can be found in ref. [8]. While $Ce_2(Pd_{0.8}Ag_{0.2})_2In$ shows a clear magnetic transition in specific heat and AC-susceptibility (left and right scales of Figure 12a), $Ce_2(Pd_{0.5}Ag_{0.5})_2In$ only shows a specific anomaly with a very weak magnetic pinch transition associated at 1 K (Figure 12b). This vanishing signal coincides with the extrapolation of $T_C(x)$ to $x = 0.5$. Notably, the specific heat of $Ce_2(Pd_{0.8}Ag_{0.2})_2In$ already show the incipient contribution of the anomaly at $T^* \sim 1$ K.

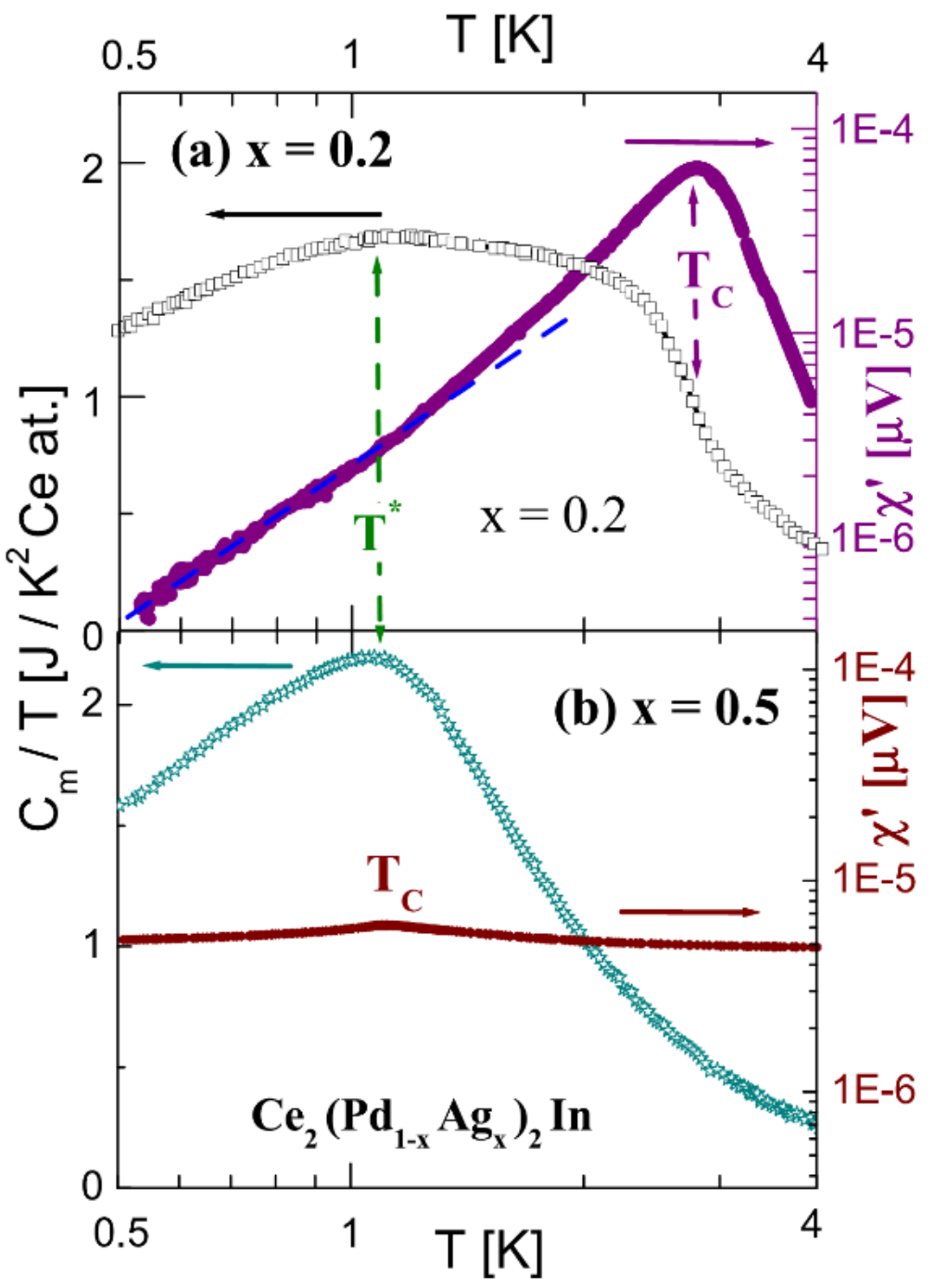

Figure 12: AC susceptibility (χ') and specific heat ($C_m$/T) (after ref. [8]) of a) of $Ce_2(Pd_{0.8}Ag_{0.2})_2In$ and b) $Ce_2(Pd_{0.5}Ag_{0.5})_2In$. Notice the logarithmic scale for χ'.

Independently, DC-magnetization (M) measurements on Ce2(Pd0.5Ag0.5)2In at 1.8 K, e.g. above *T** ~ 1 K, reveal a remanent FM contribution reflected in the *M* vs. *H* hysteresis depicted in Figure 13. This indicates the presence of two components, a weak FM and a paramagnetic one, in agreement with the NPD results that confirm the coexistence of these contributions.

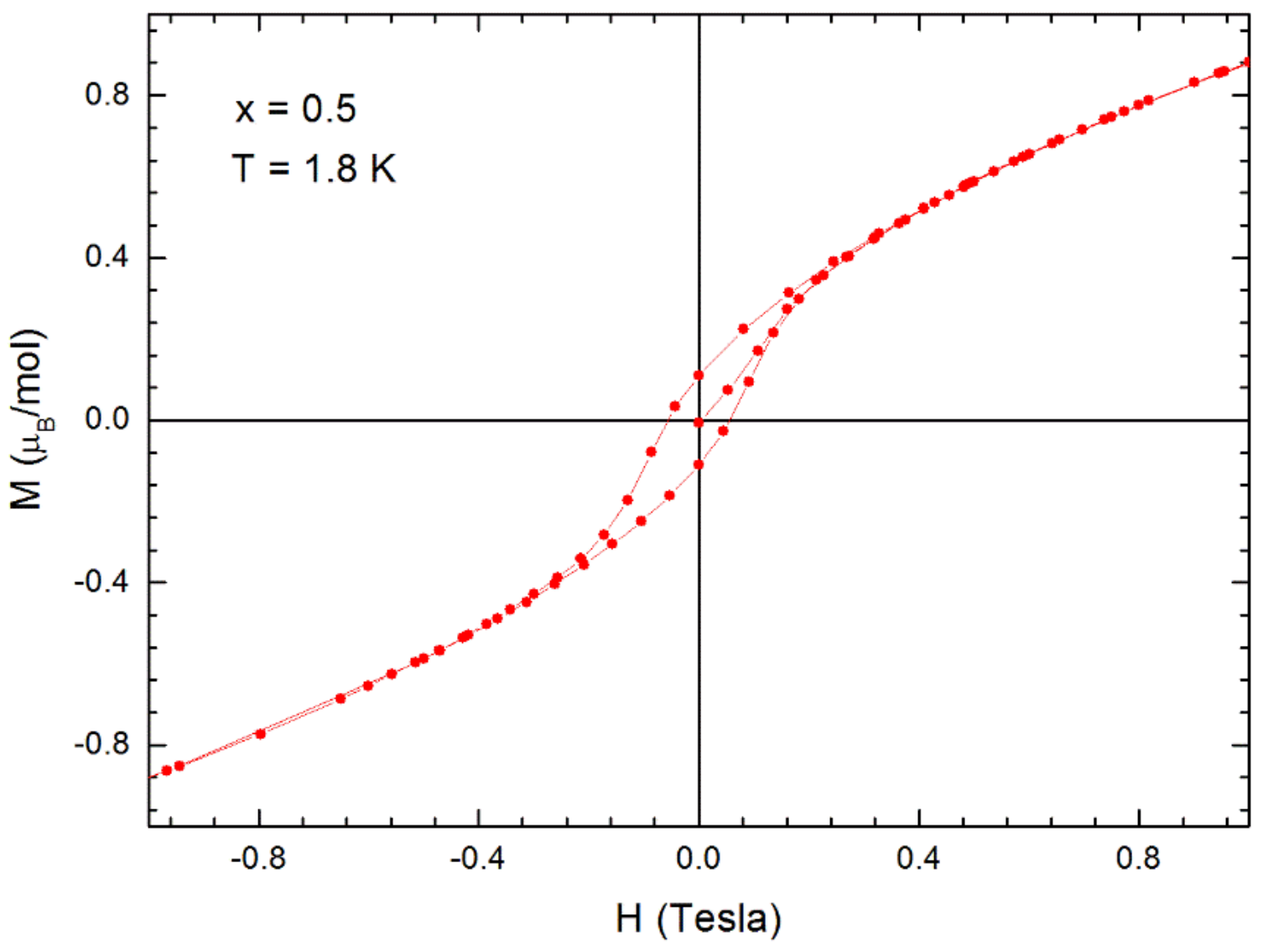

Figure 13: Remanent FM signal in $Ce_2(Pd_{0.5}Ag_{0.5})_2In$

Although specific heat measurements contained in Figure 12 clearly evidence the competition between two different components (one associated to a decreasing FM contribution and the other attributed to an arising magnetic frustration [8]), the FM signal detected by *M*(*H*) and NPD measurements cannot be associated to the former one because it exhibits a different magnetic structure as depicted in Figure 7. The origin of this weak in-plane FM arrangement is not evident, however small inhomogeneities in the atomic distribution of Pd and/or Ag may be responsible for the formation of small FM clusters involving the neighbouring Ce atoms.

According to ref. [7], the substitution of Pd with Ag is expected to increase the density of the conduction electrons and consequently to strengthen the ferromagnetism. Conversely, a weakening is experimentally observed, since in $Ce_2(Pd_{0.8}Ag_{0.2})_2In$ the ordered ferromagnetic moment is significantly reduced, compared to $Ce_2Pd_2(In_{0.6}Ce_{0.4})$, whereas in $Ce_2(Pd_{0.5}Ag_{0.5})_2In$ is almost suppressed. Nonetheless, it is worth to note that both $Ce_2Pd_2(In_{0.6}Ce_{0.4})$ and $Ce_2(Pd_{0.8}Ag_{0.2})_2In$ display the same type of ferromagnetic arrangement, with ordered moments aligned along the *c*-axis (Figure 3), whereas in $Ce_2(Pd_{0.5}Ag_{0.5})_2In$ the weak moments are ordered along the *b*-axis (Figure 7). This result suggests that: 1) Ag weakens the FM interactions; 2) favours a different kind of magnetic interaction than those taking place in $Ce_2(Pd_{0.8}Ag_{0.2})_2In$. On these bases, it can be concluded that a competition between different types of magnetic correlations arises after Ag-substitution, giving rise to a magnetically frustrated component up to a limit observed in $Ce_2(Pd_{0.5}Ag_{0.5})_2In$.

## Conclusions

Neutron powder diffraction data collected at low temperature reveal a very different magnetic ground state in $Ce_2(Pd_{0.8}Ag_{0.2})_2In$ than in $Ce_2(Pd_{0.5}Ag_{0.5})_2In$, found in the ferromagnetic branch of $Ce_2Pd_2In$. The $Ce_2(Pd_{0.8}Ag_{0.2})_2In$ compound exhibits the same ferromagnetic ordering observed in $Ce_{47}Pd_{37}In_{16}$ ($Ce_{2.22}Pd_{1.85}In_{0.78}$, see ref. [5]), with Ce moments aligned along the *c*-axis and a significant reduction of the ordered magnetic moment. With further increase of the Ag content, a different kind of ferromagnetic ordering emerges in $Ce_2(Pd_{0.5}Ag_{0.5})_2In$. Due to its extremely reduced signal, it is not clear whether this magnetic ordering is a bulk property or it is actually related to sample inhomogeneities. In any case, these results indicate that Ag-substitution induces a competition between different types of magnetic correlations. As a consequence, ferromagnetic and frustrated states coexist at the microscopic scale and the magnetically frustrated component becomes dominant at the highest level of Ag content. These observations confirm that the specific heat anomaly emerging at $T$ ~ 1 K reported in ref [8] has a completely different origin than the vanishing ferromagnetic transition.

## Acknowledgements

Authors acknowledge the 2014-2016 MAE bilateral project Italy-Argentina titled "Synthesis and study of magnetically frustrated systems with high anisotropy" for the financial support.